\documentstyle[twoside,fleqn,espcrc2]{article}
\def\beq{\begin{equation}}
\def\eeq{\end{equation}}
\def\bea{\begin{array}}
\def\eea{\end{array}}
\def\beqa{\begin{eqnarray}}
\def\eeqa{\end{eqnarray}}

\def\u1{{U(1)}}
\def\su2{{SU(2)}}
\newcommand{\re}{\relax{\rm I\kern-.18em R}}

\newcommand{\AmS}{{\protect\the\textfont2
  A\kern-.1667emt\lower.5ex\hbox{M}\kern-.125emS}}

\title{On the Doubling Phenomenon in Lattice Chern-Simons Theories
\thanks{Presented by 
P. Sodano at 
Lattice '00, Bangalore, India.
\newline
$^{\dagger}$Supported by a Swiss National Science Foundation fellowship.
}
\vskip-3cm\hfill\small DFUPG-53-00\vskip2.6cm}
\author{F. Berruto$^{(a)}$, M.C. Diamantini$^{\dagger}$$^{(b)}$ and P. Sodano$^{(a)}$
\vspace{6pt}\\ {a)Dipartimento di Fisica and Sezione I.N.F.N.,
\vskip0.1cm Universit\'a di Perugia, Via Pascoli I-06123 Perugia, Italy}
\vspace{6pt}\\ {b)CERN, Theory Division, CH-1211 Geneva, Switzerland}}
\begin{document}
\begin{abstract}
We analyse the pure Chern-Simons theory on an Euclidean infinite lattice. 
We point out that, as a consequence of its symmetries, the Chern-Simons theory
 does not have an integrable kernel. Due to the linearity of the action in 
the derivatives, the situation is very similar to the one arising in the 
lattice formulation of fermionic theories. Doubling of bosonic degrees of 
freedom is removed by adding a Maxwell term with a mechanism similar to the 
one proposed by Wilson for the fermionic theories.  
\end{abstract}
\maketitle
\section{Introduction}

In odd space-time dimensions, there is the
possibility of adding a gauge 
invariant, topological Chern-Simons (CS) term to the gauge field action. 
The CS term breaks 
both the parity and time-reversal symmetries and, when coupled with a Maxwell or
Yang-Mills term, leads to massive gauge excitations~\cite{jackiw}. 
For an Abelian model in three space-time dimensions, the pure CS 
Lagrangian is defined as
\begin{equation}
{\cal L}_{CS}=\frac{k}{2}A_{\mu}\epsilon^{\mu \alpha
\nu}\partial_{\alpha}A_{\nu}\quad ,
\label{lagrangian1}
\end{equation} 
where $k$ is a dimensionless coupling constant. 

The pure CS theory is a topological field theory~\cite{witt}. 
Being dominant at large distances, the CS
action may be used as a low energy
effective field theory for condensed matter systems\cite{zee}.

While in the continuum the pure CS theory is exactly solvable, things are quite
different on the lattice: in fact the kernel defining the CS action  exhibits 
a set of zeroes which are not due to gauge invariance~\cite{marchetti} and 
the theory is not integrable even after gauge fixing. The action 
(\ref{lagrangian1}) is of first order in the derivatives, and the
appearence of extra zeros in its lattice formulation is reminescent of the
``doubling'' of fermions on the lattice~\cite{doubling}. Doubling phenomena 
for bosonic degrees of freedom have been already evidenced in Euclidean 
lattice gravity~\cite{pelissetto} and in the analysis of the coupling of 
gravity to matter on the lattice~\cite{caselle}. 
 
Previous studies of pure CS theory on the lattice  have been carried out
using the Hamiltonian formalism in~\cite{seme}, by introducing a mixed CS 
action with two gauge fields with opposite parity or by means of two gauge 
fields living on the links of two dual lattices (thereby obtaining in both 
cases a parity even action)~\cite{kantor}.

Here we shall evidence that as a result of the theorem proved in~\cite{noi} 
the  non-integrability of the CS kernel is a general feature of 
any gauge-invariant, local, parity odd and cubic symmetric gauge theory on 
an infinite Euclidean lattice provided that under parity 
$A_{\mu}(\vec{x})\longrightarrow A_{\mu}(-\vec{x}-\hat{\mu})$ and
 $A_{\mu}(\vec{p})\longrightarrow -A_{\mu}(-\vec{p})$. Since the addition of 
a Maxwell term regularizes the CS action, the presence
of the extra 
zeros in the CS action did not cause any problem
in previous investigations of the Maxwell-CS action on the
lattice~\cite{marchetti,luscher2}. 
Although in physical applications one is naturally lead to study the
dynamics of gauge models described by a Maxwell-CS action, the proper
definition of a pure CS action on the lattice is an interesting problem in
its own right since it could open the way to a lattice evaluation of
topological invariants~\cite{witt}.

The question naturally arises if it is possible to define a parity on the 
lattice so that the Chern-Simons term is parity odd but the kernel is 
integrable. 

\section{Euclidean Chern-Simons theory on the lattice}

We consider an infinite Euclidean cubic lattice with lattice spacing $a$,
which we set to unity ($a=1$). We shall denote a lattice 
site by the vector $\vec{x}$ and a link between $\vec{x}$ and
$\vec{x}+\hat{\mu}$ ($\mu=0,1,2$) by $(\vec{x},\hat{\mu})$. 
Forward and backward difference operators are given by
$d_{\mu}f(\vec{x})=f(\vec{x}+\hat{\mu})-f(\vec{x})=
(S_{\mu}-1)f(\vec{x})$ and $\hat{d}_{\mu}f(\vec{x})
=f(\vec{x})-f(\vec{x}-\hat{\mu})=(1-S_{\mu}^{-1})f(\vec{x})$, where
$S_{\mu}f(\vec{x})=f(\vec{x}+\hat{\mu})$, $S^{-1}_{\mu}f(\vec{x})=
f(\vec{x}-\hat{\mu})$ are the forward and backward shift operators 
respectively. Summation by parts on the lattice interchanges the forward and 
backward derivatives:
$\sum_{\vec{x}}f(\vec{x})d_{\mu}g(\vec{x})
=-\sum_{\vec{x}}\hat{d}_{\mu}f(\vec{x}) g(\vec{x})$.

The lattice Fourier transformation of the 
gauge field $A_{\mu}$ is given by
\begin{equation}
A_{\mu}(\vec{x})=\int_{\cal{B}} \frac{d^3 p}{(2\pi)^3} e^{-i
\vec{p}\cdot\vec{x}} e^{-ip_{\mu}/2}A_{\mu}(\vec{p})\quad .
\label{fourier1}
\end{equation}
Due to the phase factor $e^{-ip_{\mu}/2}$, $A_{\mu}(\vec{p})$ is
antiperiodic if 
$p_{\mu}\longrightarrow p_{\mu}+2\pi (2n+1)$, with $n$ integer. 
The integration over momenta in eq.(\ref{fourier1}) 
is restricted to the Brillouin zone 
${\cal B}=\left\{p_{\mu}| -\pi\le p_{\mu}\le \pi \ ,\ \mu =0,1,2\right\}$. 
Under parity, which on an Euclidean cubic lattice corresponds to the
simultaneous
 inversion of all three directions,  
\begin{equation}
A_{\mu}(\vec{x})\longrightarrow A_{\mu}(-\vec{x}-\hat{\mu})
\end{equation}
 and
\begin{equation}
A_{\mu}(\vec{p})\longrightarrow -A_{\mu}(-\vec{p})\quad .
\end{equation}
The CS action on an Euclidean lattice derived by Fr\"{o}lich 
and Marchetti~\cite{marchetti} is:
\begin{equation}
S=\sum_{\vec{x}}A_{\mu}(\vec{x})\tilde{K}_{\mu \nu}(\vec{x} -
\vec{y})A_{\nu}(\vec{y})\quad ,
\label{action1}
\end{equation}
where $\tilde{K}_{\mu\nu}=K_{\mu\nu}+\hat{K}_{\mu\nu}$, and
\begin{eqnarray}
K_{\mu\nu}(\vec{x} -
\vec{y})&=&S_{\mu}^{\vec{y}}\ \epsilon_{\mu\alpha\nu}\ d_{\alpha}^{\vec{y}}\
\delta_{\vec{x},\vec{y}}\label{ker1}\quad ,\\
\hat{K}_{\mu\nu}(\vec{x} - \vec{y})&=&S_{\nu}^{-1,\vec{y}}\ \epsilon_{\mu
\alpha\nu}\ \hat{d}_{\alpha}^{\vec{y}}\
\delta_{\vec{x},\vec{y}}\label{ker2}\quad .
\end{eqnarray}
$K$ and $\hat{K}$ are exchanged by summation by parts. 

Although both $K$ and $\hat{K}$  define  a gauge invariant and parity odd 
kernel, Bose symmetry~\cite{karsten} requires that only the linear combination 
$\tilde{K}=K+\hat{K}$ respects this symmetry and thus provides an acceptable
definition of the lattice CS action. 
The operator $\tilde{K}(p)=K(p)+\hat{K}(p)$ has, apart from the zero
mode associated with gauge invariance, 
eigenvalues given by
\begin{equation}
\tilde{\lambda}(p)=\pm 2\sqrt{1+\cos\sum_{\mu
=0}^2p_{\mu}}\sqrt{3-\sum_{\mu=0}^2\cos p_{\mu}}\quad 
\end{equation}
and thus $\tilde{\lambda}=0$ whenever $\cos\sum_{\mu=0}^2p_{\mu}=-1$, $i.e.$
when $\sum_{\mu=0}^2p_{\mu}=
(2n+1)\pi$. The CS action (\ref{action1}) is therefore not integrable.

The properties of $K$ and $\hat{K}$ parallel the ones  of the forward and
backward derivatives, which in 
momentum space read $d_{\mu}\rightarrow e^{ip_{\mu}/2}\hat{p}_{\mu}$ and 
$\hat{d}_{\mu}\rightarrow e^{-ip_{\mu}/2}\hat{p}_{\mu}$ 
with $\hat{p}_{\mu}=2\sin p_{\mu}/2$: they do not have extra zeroes inside 
the Brillouin zone, but their linear combination 
$d + \hat{d} \rightarrow 2 \cos(p_\mu/2) \hat{p}_\mu $ has zeros at the
border of the Brillouin zone $p_\mu = \pm \pi$.

The appearance of the extra zeroes is not due to the specific form of the 
kernel in (\ref{action1}). In fact, as proven in~\cite{noi}, if the action
\begin{equation}
S=\sum_{\vec{x},\vec{y}}A_{\mu}(\vec{x})G_{\mu
\nu}(\vec{x}-\vec{y})A_{\nu}(\vec{y})\quad 
\label{action2}
\end{equation}
\noindent i) is local on the lattice;

\noindent ii) is gauge invariant: $\hat{d^x_{\mu}}G_{\mu
\nu}(\vec{x}-\vec{y})
=\hat{d^y_{\nu}}G_{\mu \nu}(\vec{x}-\vec{y})=0$;

\noindent iii) is odd under parity;

\noindent (\ref{action2}) is not integrable.

The proof of the theorem is easier in momemtum space. Bose symmetry and parity
oddness imply that the kernel $\tilde{G}_{\mu \nu}(\vec{p})
=e^{ip_{\mu}/2}G_{\mu\nu}(\vec{p})e^{-ip_{\nu}/2}$
(no sum over $\mu$ and $\nu$) is antisymmetric:
\begin{equation}
\tilde{G}_{\mu \nu}(\vec{p})=-\tilde{G}_{\nu \mu}(\vec{p})\ . 
\label{kodd}
\end{equation}
The Poincar\'e lemma~\cite{luscher} then enable us to rewrite
$\tilde{G}_{\mu \nu}
(\vec{p})$ in term of a scalar function $f(\vec{p})$: 
$\tilde{G}_{\mu \nu}(\vec{p})= 
\epsilon_{\rho \mu \nu} \hat{p}_{\rho}f(\vec{p})$.
Making use of the fact that ${G}_{\mu \nu}(\vec{p})$ must be a periodic function
of the momenta, it is easy to see that for $p_0=p_1=p_2=\pm \pi$ 
one gets
\begin{equation}
f(\pm \pi,\pm \pi,\pm \pi)=-f(\pm \pi,\pm \pi,\pm \pi)=0\quad \ .
\label{ff0}
\end{equation} 
Since the spectrum of $G_{\mu \nu}(\vec{p})$ is given by
$G(\vec{p})=\pm |f(\vec{p})|\sqrt{\sum_{\mu=0}^2 \hat{p}^2_{\mu}}$,
 eq.(\ref{ff0}) implies that the kernel 
$G(\vec{p})$ 
exhibits extra zeroes at the edges of the Brillouin zone 
and is thus not integrable.

Relaxing the assumption iii) one may study the general form of a
gauge invariant local action in three dimensions.
With the help  of the Poincar\'e lemma~\cite{luscher} it is easy to show that
the kernel $\tilde{G}_{\mu \nu}(\vec{p})$ can be divided into the sum of parity
even and parity odd terms. Since, due to locality, $\tilde{G}_{\mu
\nu}(\vec{p})$ is an
analytic function of $\vec p$, it may be expanded in Taylor series: all the
terms having even power of the momenta are parity even, while the terms with
odd power of the momenta are parity odd.
The terms with the lowest number of derivatives in this expansion are the
CS term defined in~\cite{marchetti} and the Maxwell term, whose kernel on
the lattice
is:
\begin{equation}
M_{\mu \nu}=-\Box\delta_{\mu \nu}+d_{\mu}\hat{d}_{\nu}=K_{\mu
\rho}\hat{K}_{\rho \nu} ,
\label{max}
\end{equation}
where $\Box=\sum_{\mu=0}^2d_{\mu}\hat{d}_{\mu}$ 
is the Laplacean in three dimensions.
Since all the parity odd terms fullfill the assumptions of the theorem
they generate extra zeroes in the spectrum. The only gauge invariant way to
 regularize the CS action is then the addition of a 
parity even term such as the Maxwell term. 

For the Maxwell-CS
theory on the 
lattice the kernel $\Gamma_{\mu \nu}$ may be written as 
\begin{equation}
\Gamma_{\mu \nu}=\frac{1}{4 e^2}M_{\mu \nu}
+i k G_{\mu \nu}.
\label{kmcs}
\end{equation}
In (\ref{kmcs}) $k$ is
dimensionless and $e^2$ has the dimension of a mass; the Maxwell term is an
irrelevant operator and the CS action dominates in the infrared region.
The Fourier transform of $\Gamma_{\mu \nu}$, apart from a zero mode due to
gauge 
invariance, has eigenvalues given by
\begin{equation}
\lambda_{MCS}(\vec p)=\frac{1}{2e^2}\sum_{\mu =0}^2(1-\cos p_{\mu})+
i k G(\vec p)\quad ,
\label{eigenmcs}
\end{equation}
and, as it stands, it is free from extra zeroes in the Brillouin zone since
the first term in (\ref{eigenmcs}), which is the Fourier transform 
of the Maxwell kernel, is zero only at zero momentum, and at the corners of the
Brillouin zone, $p_{\mu}=\pm \pi,\ \mu=0,1,2$, takes the value 
$\lambda_{MCS}(\vec p) =3/e^2$. 

Since the CS action is purely immaginary, the addition of the Maxwell term 
is used also in the continuum theory to provide a proper definition of the
functional
integral in the partition function of the pure CS theory. The CS limit is
reached also there
by taking the limit $e^2 \longrightarrow \infty$ after Gaussian integration.

The regularization of the extra zeros in the CS action by adding a Maxwell term
and thereby opening a gap in the fermion spectrum is similar to the 
mechanism of the Wilson fermion where a gap is opened and the energy does not
have secondary minima at the non-zero corners of the Brillouin zone.
As in the case of the Wilson fermions~\cite{kogwil}, the regularization is
done by means of an
irrelevant operator and the continuum limit $a \longrightarrow 0$ is
not changed by this addition.
Moreover, as the Wilson action explicitly breaks chiral symmetry, the action
obtained after the addition of the Maxwell term is not anymore defined under
parity.

\section{Concluding remarks}

In~\cite{noi} we pointed out a no-go theorem in the lattice regularization of 
the pure CS theory, if one requires locality, gauge invariance and parity 
oddness on the lattice. As we already pointed out in the introduction, 
a doubling phenomenon has been already advocated for bosonic theories 
on the lattice~\cite{pelissetto}~\cite{caselle}. In particular the authors 
of~\cite{pelissetto} and~\cite{caselle} found a doubling phenomenon for 
a class of lattice formulations of gravity and in the presence of a 
gravitational background all matter fields exhibit the same degeneracy as 
the graviton and the chiral fermion. This doubling phenomenon makes 
ill defined all matter fields in a gravitational background 
and it can be removed just in 
the flat space limit in the case of scalar and gauge bosonic fields. 
However, in~\cite{noi} we found a doubling phenomenon for gauge fields 
completely independent from the metric of the manifold on which the fields 
live, since the pure CS theory is a topological theory. 
Our analysis evidences explicitly that doubling 
phenomena for fields defined on a discrete space-time manifold are a very 
general phenomena due more to  the linearity of the field action 
in the derivatives than to the curvature of the manifold.    
         
The no-go theorem demonstrated in~\cite{noi} strongly relies on the 
definition of the parity transformation for the gauge field 
$A_{\mu}(\vec{x})\longrightarrow A_{\mu}(-\vec{x}-\hat{\mu})$. If one could 
define a new parity on the lattice so that the CS term is still odd but the 
kernel is integrable, then the new definition of parity should play a role 
analogous to the Ginsparg-Wilson relation~\cite{ginsparg} for lattice 
fermionic theories. For these theories the Ginsparg-Wilson relation 
hints to a new definition of chiral 
symmetry on the lattice, a generalization of the usual continuum chiral 
symmetry that is exactly recovered in the naive continuum limit.

\end{document}